\documentclass{article}
\usepackage[margin=2cm,nohead]{geometry}
\usepackage{bm,graphicx,amsbsy,amsmath,amsfonts,color,mathrsfs}
\usepackage{pgf}
\usepackage{amsthm}

\input{Qcircuit}
\newcommand{\comment}[1]{}
\newcommand{\quotes}[1]{``#1''}

\newcommand{\cnot}{\mathrm{CNOT}}
\newcommand{\ccnot}{\mathrm{CCNOT}}
\newcommand{\cswap}{\mathrm{C(SWAP)}}
\newcommand{\swap}{\mathrm{SWAP}}
\newcommand{\iswap}{\mathrm{iSWAP}}
\newcommand{\iSWAP}[1]{\mathrm{iSWAP\left(#1\right)}}
\newcommand{\PHASE}[1]{\mathrm{PHASE\left(#1\right)}}

\newcommand{\controlled}[1]{\mathrm{C}(#1)}

\newtheorem{property}{Property}[section]

\newcommand{\mat}{\left( \!\! \begin{array}{cc}}
\newcommand{\rix}{\end{array} \!\! \right)}

\def\bR{\begin{color}{red}}
\def\bB{\begin{color}{blue}}
\def\bM{\begin{color}{magenta}}
\def\bC{\begin{color}{cyan}}
\def\bW{\begin{color}{white}}
\def\bBl{\begin{color}{black}}
\def\bG{\begin{color}{green}}
\def\bY{\begin{color}{yellow}}
\def\ec{\end{color}\ }

\begin{document}


\title{Encoded Universality Of Quantum Computations \\On The Multi-Atomic Ensembles In The QED Cavity}

\author{Farid Ablayev \and Sergey Andrianov \and Sergey Moiseev \and Alexander Vasiliev}




\maketitle

\begin{abstract}
We propose an effective set of elementary quantum gates which provide an encoded universality and
demonstrate the physical feasibility of these gates for the solid-state quantum computer based on
the multi-atomic systems in the QED cavity. We use the two-qubit encoding and swapping-based
operations to simplify a physical realization of universal quantum computing and add the immunity
to a number of errors. This approach allows to implement any encoded single-qubit operation by
three elementary gates and the encoded controlled-NOT operation can be performed \emph{in a single
step}. The considerable advantages are also shown for implementing some commonly used controlled
gates.\\

\end{abstract}

\section{Introduction}
\label{sec:introduction}

During the last two decades different types of quantum computer and its physical implementations
have been
considered \cite{Nielsen-Chuang:2000:QC,Nakahara-Ohmi:2008:QC,Kok:2007:OpticalQC,Ladd:2010:QC}, where
single natural or artificial atoms, ions, molecules etc., are used for encoding of the qubits. For
these physical models a lot of universal sets of elementary unitary transformations had been
proposed \cite{Deutsch:1989:QuantumCircuits,DiVincenzo:1995:universal,Deutsch:1995:Universality,Boykin:2000:quantum-basis}.
And though there are infinitely many of them \cite{Deutsch:1995:Universality} only few sets of the
quantum gates are usually realized in experimental implementation. The most commonly used universal
set is that consisting of CNOT and single-qubit gates (e.g., see
 \cite{Nielsen-Chuang:2000:QC}). Moreover, it is well-known that even smaller discrete set
can be used to approximate any unitary operator to arbitrary accuracy. Namely, this is the set of
three one qubit gates $H$, $S$, $T$ and two qubit CNOT gate, which is usually referred to as the
\emph{standard set} \cite{Nielsen-Chuang:2000:QC}.


Physical implementation of the quantum computing on many qubits remains a main huge challenge
leading to intensive search of the novel experimental approaches. The promising approach is the one
using Heisenberg-like interactions between spin qubits. The Heisenberg interaction yields fast
two-qubit quantum gates but single-qubit gates are still the problem since they rely on weak local
interactions and hence are slow. It was
shown \cite{DiVincenzo:2000:Exchange-Interaction,Bacon:2000:Fault-Tolerant,Kempe:2001:decoherence-free-computation}
that the set of quantum gates universal for some Hilbert subspace can be built of Heisenberg-only
interactions by encoding each logical qubit by several physical qubits. This approach was termed as
\emph{encoded universality} \cite{Kempe:2001:Encoded-Universality}. This type of interaction is
universal for the case of encoding a logical qubit by at least three physical qubits. Additionally,
the proof of this fact is non-constructive and the exact sequence of elementary gates is obtained
from extensive numerical optimization \cite{Kempe:2002:Exact-gate-sequences}. The implementation of
the encoded controlled-NOT operation (up to single qubit operations) in such sequences is rather
complicated since it uses seven parallel exchange interactions or 19 serial gates.

In  \cite{Levy:2002:SpinPairs} it was also shown that controlled $\pi$-phase shift can be
achieved with Heisenberg interactions in two steps only using the encoding of logical qubits by
pairs of physical qubits. 

 Recently a new physical realization of a quantum computer has been proposed which
uses multi-atomic ensembles for encoding of single
qubits \cite{Brion:2007:Ensembles,Saffman-Molmer:2008:RydbergQC}. Multi-atomic coherent ensembles
provide a huge enhancement of the effective dipole moment that leads to a considerable acceleration
of the quantum processing rate. However, here excess excited quantum states in the multi-atomic
ensemble should be blocked in order to realize an effective two-level system providing perfect
encoding of the qubits based on the multi-atomic ensemble states. A dipole-dipole interaction is
intensively discussed for the blockade of the excess quantum states \cite{Saffman:2010:RydbergQI}.

However, the dipole blockade mechanism suffers from the decoherence problem arising due to a strong
dependence of the dipole-dipole interaction on a spatial distance between the interacting atoms.
Recently another blockade mechanism has been proposed \cite{Shahriar:2007:atomic-ensemblesQCC} based
on using a light-shift imbalance in a Raman transition. We have also proposed a novel decoherence
free blockade mechanism \cite{Moiseev:2010:multi-ensembleQC,Andrianov-Moiseev:2011:swapping-gates}
based on the collective interactions in QED cavity. Rapid development of physics and technology of
the
microcavities \cite{Duan-Kimble:2004:PhotonicQC,Aoki:2006:strong-coupling,Majer:2007:CoupledQubits}
makes this blockade mechanism a quite promising though not very simple for experimental
realization.

In this paper we propose an effective set of unitary transformations and demonstrate that logical
single and two-qubit gates can be realized naturally in the quantum computer based on the
multi-atomic ensembles in the QED cavity \cite{Andrianov-Moiseev:2011:swapping-gates}. We
\emph{explicitly} show that this set possesses an encoded universality when using the 2-qubit
encoding mentioned in \cite{Palma:1996:quantum-dissipation}: $\ket{0_L}=\ket{01}$,
$\ket{1_L}=\ket{10}$. In \cite{Byrd-Lidar:2002:Problems-of-Decoherence} it was shown that
this encoding allows to solve two major problems of solid-state quantum computing. First of all, it
eliminates the problems with implementation of single-qubit operations.
Additionally, this encoding forms a decoherence-free subspace
(DFS) \cite{Bacon:2000:Fault-Tolerant}, which allows to prevent a number of computational
errors \cite{Byrd-Lidar:2002:Problems-of-Decoherence}. In the physical model we consider
here \cite{Andrianov-Moiseev:2011:swapping-gates}, there is a specific type of error, which comes
out when the swapping operations are applied to the basis state $\ket{11}$. This error can be
suppressed by using collective blockade, which slows the computation. On the other hand, it is
obvious that the qubit encoding we use here has the immunity to this type of error.




The main advantage of our approach is the ability to perform the
controlled-NOT operation \emph{in a single step}. This is achieved by additional nonlinear frequency shift
naturally arising in the QED cavity with Heisenberg-type interaction. We also show that the proposed set of quantum
gates is efficient for implementing complex controlled operations, which are at the heart of many
efficient quantum algorithms (e.g. creating fingerprints for the technique of quantum
fingerprinting \cite{ablayev-vasiliev:2009:EPTCS}).

\section{Quantum Computer Based on Multi-Atomic Ensembles in Resonator}
\label{sec:swap}

We discuss a novel architecture of quantum computer based on the multi-atomic ensembles of two
level atoms in QED
cavity \cite{Moiseev:2010:multi-ensembleQC,Andrianov-Moiseev:2011:swapping-gates}. Here, we can
introduce the following collective states in $m$-th processing node:
\begin{equation}
\label{eq1}
\left| 0 \right\rangle _m =\left| g \right\rangle _1 \left| g \right\rangle
_2 ...\left| g \right\rangle _{N_m },
\end{equation}
corresponds to the ground state of the node,
\begin{equation}
\label{eq2}
\left| 1 \right\rangle _m =\sqrt {1 \mathord{\left/ {\vphantom {1 {N_m }}}
\right. \kern-\nulldelimiterspace} {N_m }} \sum\limits_j^{N_m } {\left| g
\right\rangle _1 \left| g \right\rangle _2 ...\left| e \right\rangle _j
...\left| g \right\rangle _{N_m } },
\end{equation}
and
\begin{equation}
\label{eq3}
\left| 2 \right\rangle _m =\sqrt {2 \mathord{\left/ {\vphantom {2 {N_m
\left( {N_m -1} \right)}}} \right. \kern-\nulldelimiterspace} {N_m \left(
{N_m -1} \right)}} \sum\limits_{i\ne j}^{N_m } {\left| g \right\rangle _1
\left| g \right\rangle _2 ...\left| e \right\rangle _j ...\left| e
\right\rangle _j ...\left| g \right\rangle _{N_m } },
\end{equation}
are the collective states of m-th node with single and two atomic
excitations (where $\left| g \right\rangle _j $ and $\left| e \right\rangle
_j $are the ground and excited states of j-th atom and $N_m $ is a number of
atoms in the m-th node).

We denote the collective state of a pair of the processing nodes 1 and 2 by
$\left| {00} \right\rangle =\left| 0 \right\rangle _1 \left| 0 \right\rangle
_2 $. Similarly, we introduce the states $\left| {01} \right\rangle $,
$\left| {10} \right\rangle $, $\left| {11} \right\rangle $, $\left| {02}
\right\rangle $, $\left| {20} \right\rangle $.
The proposed architecture provides the means of performing the following
quantum gates:

\begin{equation}
\iswap = \left(
\begin{array}{cccc}
1&0&0&0\\
0&0 & i&0\\
0&i & 0&0\\
0&0&0&1\\
\end{array}\right),
\end{equation}

\begin{equation}
\cswap = \left(
\begin{array}{cccccccc}
1&0&0&0&0&0&0&0\\
0&1&0&0&0&0&0&0\\
0&0&1&0&0&0&0&0\\
0&0&0&1&0&0&0&0\\
0&0&0&0&1&0&0&0\\
0&0&0&0&0&0 & 1&0\\
0&0&0&0&0&1 & 0&0\\
0&0&0&0&0&0&0&1\\
\end{array}\right).
\end{equation}

\noindent It follows also from \cite{Moiseev:2010:multi-ensembleQC,Andrianov-Moiseev:2011:swapping-gates} that, using
appropriate times, it is possible to perform a gate which generalizes the iSWAP gate. Specifically,
this gate is

\begin{equation}
\iSWAP{\theta}=\left(
\begin{array}{cccc}
1&0&0&0\\
0&\cos{\frac{\theta}{2}} & i\sin{\frac{\theta}{2}}&0\\
0&i\sin{\frac{\theta}{2}} & \cos{\frac{\theta}{2}}&0\\
0&0&0&1\\
\end{array}\right)
\end{equation}
for arbitrary (experimentally possible) $\theta$. When $\theta = \pi$, this gate is exactly iSWAP.

The gate we denote here as $\iSWAP{\theta}$ describes an anisotropic exchange interaction and is
known to be universal in an encoded universality setting \cite{Kempe:2001:Encoded-Universality}.
However, we include it in a larger universal set for the reasons mentioned earlier.

The proposed architecture also allows us to perform the following gate:
\begin{equation}
\PHASE{\theta} = \left(
\begin{array}{cccc}
1&0&0&0\\
0&e^{-i\theta/2+i\phi/2} & 0&0\\
0&0 & e^{i\theta/2+i\phi/2}&0\\
0&0&0&e^{i\phi}\\
\end{array}\right),
\end{equation}
where the phases $\theta $ and $\varphi $ can be controlled by the different values of the external
magnetic (or electric) fields on the spatially distinct nodes that provides, respectively,
different Zeeman (or Stark) frequency shifts for the two-level atoms localized in the
nodes \cite{Moiseev:2010:multi-ensembleQC}.

The implementation of universal quantum computations based on the set of iSWAP and single qubit
gates in the proposed model meets some obstacles:

- while the two-qubit gates can be implemented quite quickly, the
single-qubit gates require blockade, which leads to a significant loss of
performance;

- the application of the ``swapping'' gates $\iSWAP{\theta}$ and $\cswap$ to the basis state
$\ket{11}$ turns it into $(\ket{02} + \ket{20})/\sqrt{2}$ and backwards. This side effect can be
avoided by using collective blockage, which tunes down the performance as well.

\subsection{Implementing universal quantum computations with $\cswap$, $\iSWAP{\theta}$,
$\PHASE{\theta}$} \label{section:Universality}

In this section we present an approach for implementing universal quantum computations using the
set of elementary quantum gates available in the considered physical model of a quantum computer.
At the heart of our approach lies the idea of using logical qubits encoded by two physical qubits.
This idea was proposed in \cite{Palma:1996:quantum-dissipation}.

Such encoding implies that we consider pairs of processing nodes in states $\ket{01}$, $\ket{10}$
as single logical qubits in states $\ket{0_L}$, $\ket{1_L}$ respectively. This encoding is somewhat
similar to dual rail logic in classical computations for it adds robustness to the system forming a
decoherence-free subspace \cite{Bacon:2000:Fault-Tolerant,Byrd-Lidar:2002:Problems-of-Decoherence}.

\begin{figure}
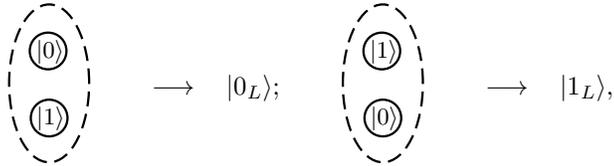

\begin{center}
\begin{tabular}{rclrcl}
\begin{tabular}{r}
\begin{pgfpicture}{21.67mm}{67.71mm}{36.32mm}{92.72mm}
\pgfsetxvec{\pgfpoint{1.00mm}{0mm}}
\pgfsetyvec{\pgfpoint{0mm}{1.00mm}}
\color[rgb]{0,0,0}\pgfsetlinewidth{0.30mm}\pgfsetdash{}{0mm}
\pgfcircle[stroke]{\pgfxy(28.99,75.61)}{2.44mm}
\pgfcircle[stroke]{\pgfxy(28.85,84.53)}{2.44mm}
\pgfsetdash{{2.00mm}{1.00mm}}{0mm}\pgfellipse[stroke]{\pgfxy(28.99,80.22)}{\pgfxy(5.32,0.00)}{\pgfxy(0.00,10.50)}
\pgfputat{\pgfxy(27.24,83.60)}{\pgfbox[bottom,left]{\fontsize{8.54}{10.24}\selectfont $\ket{0}$}}
\pgfputat{\pgfxy(27.32,74.73)}{\pgfbox[bottom,left]{\fontsize{8.54}{10.24}\selectfont $\ket{1}$}}
\end{pgfpicture}
\end{tabular} &
$\longrightarrow$ &
$\ket{0_L}$; &
\begin{tabular}{r}
\begin{pgfpicture}{21.67mm}{67.71mm}{36.32mm}{92.72mm}
\pgfsetxvec{\pgfpoint{1.00mm}{0mm}}
\pgfsetyvec{\pgfpoint{0mm}{1.00mm}}
\color[rgb]{0,0,0}\pgfsetlinewidth{0.30mm}\pgfsetdash{}{0mm}
\pgfcircle[stroke]{\pgfxy(28.99,75.61)}{2.44mm}
\pgfcircle[stroke]{\pgfxy(28.85,84.53)}{2.44mm}
\pgfsetdash{{2.00mm}{1.00mm}}{0mm}\pgfellipse[stroke]{\pgfxy(28.99,80.22)}{\pgfxy(5.32,0.00)}{\pgfxy(0.00,10.50)}
\pgfputat{\pgfxy(27.24,83.60)}{\pgfbox[bottom,left]{\fontsize{8.54}{10.24}\selectfont $\ket{1}$}}
\pgfputat{\pgfxy(27.32,74.73)}{\pgfbox[bottom,left]{\fontsize{8.54}{10.24}\selectfont $\ket{0}$}}
\end{pgfpicture}
\end{tabular} &
$\longrightarrow$ &
$\ket{1_L}$, \\
\end{tabular}
\end{center}
\vspace*{8pt}
  \caption{Pairwise qubit encoding. Small circles denote the processing nodes in the indicated quantum states.}
  \label{PairwiseEncoding}
\end{figure}

In this case any single qubit state $\alpha\ket{0_L}+\beta\ket{1_L}$ is actually stored as an
entangled two qubit state $\alpha\ket{01}+\beta\ket{10}$, that is the basis state of such a
composite qubit is determined by the basis state of the first qubit of a pair.

Note also, that this encoding excludes the usage of the single qubit operations, plus it prohibits
the pairs of processing nodes to be in state $\ket{11}$. Hence it solves both of physical problems
stated earlier.

The following property explicitly shows that in such encoding of qubits the two and three qubit
operations available in our physical model are sufficient to implement the standard set, thus
proving their encoded universality.

\begin{property}
The set of quantum gates $\{\cswap, \iSWAP{\theta}, \PHASE{\theta}\}$ is universal for the
Hilbert subspace spanned by encoded states $\ket{0_L}=\ket{01}$, $\ket{1_L}=\ket{10}$.
\end{property}
\begin{proof}
First of all we show the effect of our elementary quantum gates when acting on pairs of nodes in
basis states $\ket{01}$, $\ket{10}$ and their linear combinations.

For instance, the $\swap$ operation turns $\ket{01}$ into $\ket{10}$ and backwards, thus acting on
a pair like the gate $X$ (the NOT gate):

\begin{figure}
$$\begin{array}{ccc}
\begin{array}{l}
\begin{pgfpicture}{0.23mm}{67.71mm}{36.32mm}{92.72mm}
\pgfsetxvec{\pgfpoint{1.00mm}{0mm}}
\pgfsetyvec{\pgfpoint{0mm}{1.00mm}}
\color[rgb]{0,0,0}\pgfsetlinewidth{0.30mm}\pgfsetdash{}{0mm}
\pgfcircle[stroke]{\pgfxy(28.99,75.61)}{2.44mm}
\pgfcircle[stroke]{\pgfxy(28.85,84.53)}{2.44mm}
\pgfsetdash{{2.00mm}{1.00mm}}{0mm}\pgfellipse[stroke]{\pgfxy(28.99,80.22)}{\pgfxy(5.32,0.00)}{\pgfxy(0.00,10.50)}
\pgfputat{\pgfxy(3.09,79.21)}{\pgfbox[bottom,left]{\fontsize{14.23}{17.07}\selectfont SWAP}}
\pgfsetdash{}{0mm}\pgfmoveto{\pgfxy(2.23,78.20)}\pgflineto{\pgfxy(17.77,78.20)}\pgflineto{\pgfxy(17.77,83.67)}\pgflineto{\pgfxy(2.23,83.67)}\pgfclosepath\pgfstroke
\pgfmoveto{\pgfxy(17.77,83.81)}\pgfcurveto{\pgfxy(18.66,85.20)}{\pgfxy(20.01,86.23)}{\pgfxy(21.59,86.72)}\pgfcurveto{\pgfxy(22.79,87.10)}{\pgfxy(24.08,87.14)}{\pgfxy(25.30,86.81)}\pgfcurveto{\pgfxy(26.65,86.45)}{\pgfxy(27.85,85.65)}{\pgfxy(28.70,84.53)}\pgfstroke
\pgfmoveto{\pgfxy(28.70,84.53)}\pgflineto{\pgfxy(26.84,86.74)}\pgflineto{\pgfxy(27.00,85.76)}\pgflineto{\pgfxy(26.03,85.60)}\pgflineto{\pgfxy(28.70,84.53)}\pgfclosepath\pgffill
\pgfmoveto{\pgfxy(28.70,84.53)}\pgflineto{\pgfxy(26.84,86.74)}\pgflineto{\pgfxy(27.00,85.76)}\pgflineto{\pgfxy(26.03,85.60)}\pgflineto{\pgfxy(28.70,84.53)}\pgfclosepath\pgfstroke
\pgfmoveto{\pgfxy(17.54,78.07)}\pgfcurveto{\pgfxy(17.88,76.12)}{\pgfxy(19.25,74.50)}{\pgfxy(21.12,73.84)}\pgfcurveto{\pgfxy(22.24,73.44)}{\pgfxy(23.46,73.43)}{\pgfxy(24.64,73.65)}\pgfcurveto{\pgfxy(26.19,73.95)}{\pgfxy(27.65,74.62)}{\pgfxy(28.87,75.63)}\pgfstroke
\pgfmoveto{\pgfxy(28.87,75.63)}\pgflineto{\pgfxy(26.07,74.93)}\pgflineto{\pgfxy(27.02,74.64)}\pgflineto{\pgfxy(26.73,73.69)}\pgflineto{\pgfxy(28.87,75.63)}\pgfclosepath\pgffill
\pgfmoveto{\pgfxy(28.87,75.63)}\pgflineto{\pgfxy(26.07,74.93)}\pgflineto{\pgfxy(27.02,74.64)}\pgflineto{\pgfxy(26.73,73.69)}\pgflineto{\pgfxy(28.87,75.63)}\pgfclosepath\pgfstroke
\end{pgfpicture}
\end{array}
& \longrightarrow
&
\begin{array}{l}
\quad \Qcircuit @C=1.0em
@R=1.0em {
 & \gate{X} & \qw\\
}
\end{array}
\end{array}
$$
\vspace*{8pt}
  \caption{Logical NOT gate implemented by the physical SWAP gate.}
  \label{SWAP-NOT-figure}
\end{figure}
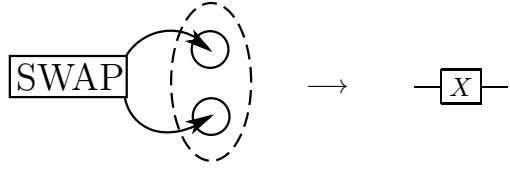

In the similar manner $\cswap$ actually implements the $\cnot$ gate:
\begin{figure}
$$\begin{array}{ccc}
\begin{array}{l}
\begin{pgfpicture}{-6.45mm}{67.71mm}{50.79mm}{97.18mm}
\pgfsetxvec{\pgfpoint{1.00mm}{0mm}}
\pgfsetyvec{\pgfpoint{0mm}{1.00mm}}
\color[rgb]{0,0,0}\pgfsetlinewidth{0.30mm}\pgfsetdash{}{0mm}
\pgfcircle[stroke]{\pgfxy(28.99,75.61)}{2.44mm}
\pgfcircle[stroke]{\pgfxy(28.85,84.53)}{2.44mm}
\pgfsetdash{{2.00mm}{1.00mm}}{0mm}\pgfellipse[stroke]{\pgfxy(28.99,80.22)}{\pgfxy(5.32,0.00)}{\pgfxy(0.00,10.50)}
\pgfputat{\pgfxy(-3.88,79.50)}{\pgfbox[bottom,left]{\fontsize{14.23}{17.07}\selectfont C(SWAP)}}
\pgfsetdash{}{0mm}\pgfmoveto{\pgfxy(-4.45,78.19)}\pgflineto{\pgfxy(17.77,78.19)}\pgflineto{\pgfxy(17.77,83.67)}\pgflineto{\pgfxy(-4.45,83.67)}\pgfclosepath\pgfstroke
\pgfmoveto{\pgfxy(17.77,83.81)}\pgfcurveto{\pgfxy(18.66,85.20)}{\pgfxy(20.01,86.23)}{\pgfxy(21.59,86.72)}\pgfcurveto{\pgfxy(22.79,87.10)}{\pgfxy(24.08,87.14)}{\pgfxy(25.30,86.81)}\pgfcurveto{\pgfxy(26.65,86.45)}{\pgfxy(27.85,85.65)}{\pgfxy(28.70,84.53)}\pgfstroke
\pgfmoveto{\pgfxy(28.70,84.53)}\pgflineto{\pgfxy(26.84,86.74)}\pgflineto{\pgfxy(27.00,85.76)}\pgflineto{\pgfxy(26.03,85.60)}\pgflineto{\pgfxy(28.70,84.53)}\pgfclosepath\pgffill
\pgfmoveto{\pgfxy(28.70,84.53)}\pgflineto{\pgfxy(26.84,86.74)}\pgflineto{\pgfxy(27.00,85.76)}\pgflineto{\pgfxy(26.03,85.60)}\pgflineto{\pgfxy(28.70,84.53)}\pgfclosepath\pgfstroke
\pgfmoveto{\pgfxy(17.54,78.07)}\pgfcurveto{\pgfxy(17.88,76.12)}{\pgfxy(19.25,74.50)}{\pgfxy(21.12,73.84)}\pgfcurveto{\pgfxy(22.24,73.44)}{\pgfxy(23.46,73.43)}{\pgfxy(24.64,73.65)}\pgfcurveto{\pgfxy(26.19,73.95)}{\pgfxy(27.65,74.62)}{\pgfxy(28.87,75.63)}\pgfstroke
\pgfmoveto{\pgfxy(28.87,75.63)}\pgflineto{\pgfxy(26.07,74.93)}\pgflineto{\pgfxy(27.02,74.64)}\pgflineto{\pgfxy(26.73,73.69)}\pgflineto{\pgfxy(28.87,75.63)}\pgfclosepath\pgffill
\pgfmoveto{\pgfxy(28.87,75.63)}\pgflineto{\pgfxy(26.07,74.93)}\pgflineto{\pgfxy(27.02,74.64)}\pgflineto{\pgfxy(26.73,73.69)}\pgflineto{\pgfxy(28.87,75.63)}\pgfclosepath\pgfstroke
\color[rgb]{0,0,0}\pgfcircle[stroke]{\pgfxy(43.47,75.61)}{2.44mm}
\pgfcircle[stroke]{\pgfxy(43.32,84.53)}{2.44mm}
\pgfsetdash{{2.00mm}{1.00mm}}{0mm}\pgfellipse[stroke]{\pgfxy(43.47,80.22)}{\pgfxy(5.32,0.00)}{\pgfxy(0.00,10.50)}
\pgfsetdash{}{0mm}\pgfmoveto{\pgfxy(6.44,84.01)}\pgfcurveto{\pgfxy(10.88,92.16)}{\pgfxy(20.06,96.52)}{\pgfxy(29.18,94.81)}\pgfcurveto{\pgfxy(35.25,93.68)}{\pgfxy(40.49,89.90)}{\pgfxy(43.48,84.50)}\pgfstroke
\pgfmoveto{\pgfxy(6.44,84.01)}\pgflineto{\pgfxy(8.40,86.13)}\pgflineto{\pgfxy(7.45,85.85)}\pgflineto{\pgfxy(7.17,86.80)}\pgflineto{\pgfxy(6.44,84.01)}\pgfclosepath\pgffill
\pgfmoveto{\pgfxy(6.44,84.01)}\pgflineto{\pgfxy(8.40,86.13)}\pgflineto{\pgfxy(7.45,85.85)}\pgflineto{\pgfxy(7.17,86.80)}\pgflineto{\pgfxy(6.44,84.01)}\pgfclosepath\pgfstroke
\color[rgb]{0,0,0}\pgfputat{\pgfxy(43.42,79.50)}{\pgfbox[bottom,left]{\fontsize{5.69}{6.83}\selectfont \makebox[0pt]{control}}}
\pgfputat{\pgfxy(28.96,79.46)}{\pgfbox[bottom,left]{\fontsize{5.69}{6.83}\selectfont \makebox[0pt]{target}}}
\end{pgfpicture}
\end{array}
& \longrightarrow
&
\begin{array}{l}
\quad \Qcircuit @C=1.0em
@R=1.0em {
 & \ctrl{1} & \qw\\
 & \targ & \qw\\
}
\end{array}
\end{array}
$$
\vspace*{8pt}
  \caption{Logical $\cnot$ gate implemented by the physical $\cswap$ gate.}
  \label{CSWAP-CNOT-figure}
\end{figure}

More formally:
\begin{equation}
\begin{array}{l}
\cnot_L\left(\alpha_1\ket{0_L}\ket{0_L}+\alpha_2\ket{0_L}\ket{1_L}+\alpha_3\ket{1_L}\ket{0_L}+\alpha_4\ket{1_L}\ket{1_L}\right)=\\
\cswap\left(\alpha_1\ket{01}\ket{01}+\alpha_2\ket{01}\ket{10}+\alpha_3\ket{10}\ket{01}+\alpha_4\ket{10}\ket{10}\right)=\\
\alpha_1\ket{01}\ket{01}+\alpha_2\ket{01}\ket{10}+\alpha_3\ket{10}\ket{10}+\alpha_4\ket{10}\ket{01}=\\
\alpha_1\ket{0_L}\ket{0_L}+\alpha_2\ket{0_L}\ket{1_L}+\alpha_3\ket{1_L}\ket{1_L}+\alpha_4\ket{1_L}\ket{0_L}.
\end{array}
\end{equation}

The physical implementation of these gates based on the controlled interaction between the
ensembles of the two-level atoms and field cavity modes is described in \cite{Andrianov-Moiseev:2011:swapping-gates}.

If we look at the matrix for the generalized $\iSWAP{\theta}$ gate, it's middle part (responsible
for transforming $\ket{01}$ and $\ket{10}$ basis states) is actually a rotation by the angle
$-\theta$ about the $\hat{x}$ axis of the Bloch sphere, i.e. $\iSWAP{\theta}$ corresponds to the
following operator:
\begin{equation}
\begin{array}{lcr}
\iSWAP{\theta}=
\left(\begin{array}{l}
\Qcircuit @C=1.0em @R=1.0em {
1&0&0&0\\
0&\push{\cos{\frac{\theta}{2}}} & \push{i\sin{\frac{\theta}{2}}}&0\\
0&\push{i\sin{\frac{\theta}{2}}} & \push{\cos{\frac{\theta}{2}}}&0\\
\push{0}&0&0&\push{1}\gategroup{2}{2}{3}{3}{.7em}{--}
}
\end{array}\right)
& \longrightarrow
&
R_{\hat{x}}(\theta)=\left(
\begin{array}{cc}
\cos\frac{\theta}{2} & -i\sin\frac{\theta}{2}\\
-i\sin\frac{\theta}{2} & \cos\frac{\theta}{2}
\end{array}\right).
\end{array}
\end{equation}

Similarly, $\PHASE{\theta}$ turns our composite qubit around the axis $\hat{z}$ (up to the phase
factor of $e^{i\phi/2}$):
\begin{equation}
\begin{array}{lcccr}
\PHASE{\theta} =
e^{i\frac{\phi}{2}}\left(\begin{array}{l}
\Qcircuit @C=1.0em @R=1.0em {
\push{e^{-i\frac{\phi}{2}}}&0&0&0\\
0&\push{e^{-i\frac{\theta}{2}}} & \push{0}&0\\
0&\push{0} & \push{e^{i\frac{\theta}{2}}}&0\\
\push{0}&0&0&\push{e^{i\frac{\phi}{2}}}\gategroup{2}{2}{3}{3}{.7em}{--}
}
\end{array}\right)
& & \longrightarrow &
&
R_{\hat{z}}(\theta)=\left(
\begin{array}{cc}
e^{-{i\frac{\theta}{2}}} & 0\\
0 & e^{i\frac{\theta}{2}}
\end{array}\right).
\end{array}
\end{equation}

Since an arbitrary rotation of a single qubit (and thus any single qubit gate) can be decomposed
into the product of three rotations about orthogonal axes (say, $R_{\hat{x}}$ and $R_{\hat{z}}$),
our basis allows to avoid using operations of single processing nodes and thus blockage. All of the
single qubit gates are performed by the means two node operations.


For instance, the Hadamard transform can be implemented as follows:
\begin{equation}
H=e^{i\pi/2}R_{\hat{z}}\left(\frac{\pi}{2}\right)R_{\hat{x}}\left(\frac{\pi}{2}\right)R_{\hat{z}}\left(\frac{\pi}{2}\right)
.
\end{equation}

The other two single qubit gates $S$ and $T$ from the standard set up to the phase factor are
rotations about $\hat{z}$ axis:
\begin{equation}
S=\left(\begin{array}{cc}1 & 0\\0 & i\end{array}\right)
=e^{i\pi/4}\left(\begin{array}{cc}e^{-i\pi/4} & 0\\0 & e^{i\pi/4}\end{array}\right)
=e^{i\pi/4}R_{\hat{z}}\left(\frac{\pi}{2}\right)
\end{equation}
\begin{equation}
T=\left(\begin{array}{cc}1 & 0\\0 & e^{i\pi/4}\end{array}\right)=
e^{i\pi/8}\left(\begin{array}{cc}e^{-i\pi/8} & 0\\0 &
e^{i\pi/8}\end{array}\right)
=e^{i\pi/8}R_{\hat{z}}\left(\frac{\pi}{4}\right)
\end{equation}

Therefore, the set of gates $\{\cswap, \iSWAP{\theta}, \PHASE{\theta}\}$ allows to implement the
standard set of quantum gates, which proves it's encoded universality.

\end{proof}

Note, that even though the $\swap$ is not in our basis set it can be simulated by the $\iswap =
\iswap(\pi)$ gate up to unimportant global phase $e^{i\pi/2}$ when acting on basis states
$\ket{01}$ and
$\ket{10}$. 

Note also, that the universality of the proposed set of elementary gates rely on the presence of
\quotes{continuous} operations $\iSWAP{\theta}$ and $\PHASE{\theta}$. This fact requires a higher
precision of the hardware, but excludes the approximation algorithms for implementing arbitrary
single qubit operations using the standard set of $\cnot$, $H$, $S$, and $T$. Conversely, we may
restrict ourselves with using only $\iSWAP{\frac{\pi}{2}}$, $\PHASE{\frac{\pi}{2}}$, and
$\PHASE{\frac{\pi}{4}}$ (which is enough for implementing gates $H$, $S$, and $T$) and exploit
standard approximation schemes for arbitrary single qubit gates.

It is also well-known \cite{Nielsen-Chuang:2000:QC} that the $\cswap$ (Fredkin) gate is universal
for classical computations, since it can be used to perform logical NOT, AND, and FANOUT
operations:
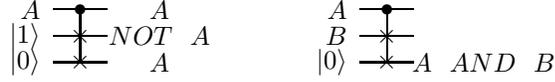
\begin{figure}
$$
\begin{array}{lcr}
\begin{array}{l}\quad~ \Qcircuit @C=1.0em @R=1.0em {
\lstick{A} & \ctrl{1} & \qw & \quad\quad{A}\\
\lstick{\ket{1}} & \qswap\qwx[1] & \qw & \quad\quad{NOT~~A}\\
\lstick{\ket{0}} & \qswap & \qw & \quad\quad{A}\\
}\end{array}
&\hspace{1.5cm}
&
\begin{array}{l}\quad~ \Qcircuit
@C=1.0em @R=1.0em {
\lstick{A} & \ctrl{1} & \qw\\
\lstick{B} & \qswap\qwx[1] & \qw\\
\lstick{\ket{0}} & \qswap & \qw & \quad\quad\quad~{A~~AND~~B}\\
}\end{array}
\end{array}
$$
\vspace*{8pt}
  \caption{Circuits performing logical NOT, AND, and FANOUT
operations.}
  \label{Figure:Fredkin-gate-classical}
\end{figure}

\subsection{Universal set based on controlled phase gate}

We have already mentioned that there is another popular approach to constructing universal quantum
computations based on the Controlled Phase gate:
\begin{equation}
\begin{array}{lcr}
\begin{array}{l}\quad~ \Qcircuit @C=1.0em @R=1.0em {
 & \ctrl{1} & \qw\\
 & \gate{e^{-i\phi}} & \qw\\
}\end{array}
& = &
\left(
\begin{array}{cccc}
1&0&0&0\\
0&1 & 0&0\\
0&0 & 1&0\\
0&0&0&e^{-i\phi}\\
\end{array}\right).
\end{array}
\end{equation}

When $\phi=\pi$ this gate becomes the Controlled-$Z$ gate
\begin{equation}
\begin{array}{lcr}
\begin{array}{l}\quad~ \Qcircuit @C=1.0em @R=1.0em {
 & \ctrl{1} & \qw\\
 & \gate{Z} & \qw\\
}\end{array}
& = C(Z) = &
\left(
\begin{array}{cccc}
1&0&0&0\\
0&1 & 0&0\\
0&0 & 1&0\\
0&0&0&-1\\
\end{array}\right),
\end{array}
\end{equation}
which can be used to construct the $\cnot$ gate:
\begin{equation}
\begin{array}{lr}
\begin{array}{l}\quad~ \Qcircuit @C=1.0em @R=1.0em {
 & \ctrl{2} & \qw\\
 \\
 & \targ & \qw\\
}\end{array}
=
\begin{array}{l}\Qcircuit @C=1.0em @R=1.0em {
 & \qw & \ctrl{2} & \qw & \qw\\
 \\
 & \gate{H} & \gate{Z} & \gate{H} & \qw\\
}\end{array}
\end{array}
\end{equation}

It can be easily verified that we can implement the Controlled Phase gate in our pairwise qubit
encoding by simply applying it to the first processing nodes of each pair:
\begin{figure}
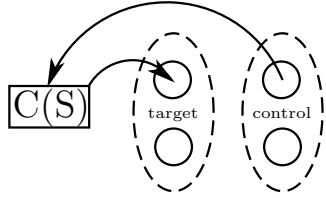

\begin{center}
\begin{tabular}{c}
\begin{pgfpicture}{5.16mm}{67.71mm}{50.79mm}{96.85mm}
\pgfsetxvec{\pgfpoint{1.00mm}{0mm}}
\pgfsetyvec{\pgfpoint{0mm}{1.00mm}}
\color[rgb]{0,0,0}\pgfsetlinewidth{0.30mm}\pgfsetdash{}{0mm}
\pgfcircle[stroke]{\pgfxy(28.99,75.61)}{2.44mm}
\pgfcircle[stroke]{\pgfxy(28.85,84.53)}{2.44mm}
\pgfsetdash{{2.00mm}{1.00mm}}{0mm}\pgfellipse[stroke]{\pgfxy(28.99,80.22)}{\pgfxy(5.32,0.00)}{\pgfxy(0.00,10.50)}
\pgfputat{\pgfxy(7.64,79.55)}{\pgfbox[bottom,left]{\fontsize{14.23}{17.07}\selectfont C(S)}}
\pgfsetdash{}{0mm}\pgfmoveto{\pgfxy(7.16,78.19)}\pgflineto{\pgfxy(17.77,78.19)}\pgflineto{\pgfxy(17.77,83.71)}\pgflineto{\pgfxy(7.16,83.71)}\pgfclosepath\pgfstroke
\pgfmoveto{\pgfxy(17.77,83.81)}\pgfcurveto{\pgfxy(18.78,85.52)}{\pgfxy(20.48,86.70)}{\pgfxy(22.42,87.04)}\pgfcurveto{\pgfxy(23.18,87.18)}{\pgfxy(23.95,87.18)}{\pgfxy(24.70,87.04)}\pgfcurveto{\pgfxy(26.31,86.75)}{\pgfxy(27.74,85.85)}{\pgfxy(28.70,84.53)}\pgfstroke
\pgfmoveto{\pgfxy(28.70,84.53)}\pgflineto{\pgfxy(26.90,86.78)}\pgflineto{\pgfxy(27.03,85.80)}\pgflineto{\pgfxy(26.05,85.67)}\pgflineto{\pgfxy(28.70,84.53)}\pgfclosepath\pgffill
\pgfmoveto{\pgfxy(28.70,84.53)}\pgflineto{\pgfxy(26.90,86.78)}\pgflineto{\pgfxy(27.03,85.80)}\pgflineto{\pgfxy(26.05,85.67)}\pgflineto{\pgfxy(28.70,84.53)}\pgfclosepath\pgfstroke
\color[rgb]{0,0,0}\pgfcircle[stroke]{\pgfxy(43.47,75.61)}{2.44mm}
\pgfcircle[stroke]{\pgfxy(43.32,84.53)}{2.44mm}
\pgfsetdash{{2.00mm}{1.00mm}}{0mm}\pgfellipse[stroke]{\pgfxy(43.47,80.22)}{\pgfxy(5.32,0.00)}{\pgfxy(0.00,10.50)}
\pgfsetdash{}{0mm}\pgfmoveto{\pgfxy(12.42,84.09)}\pgfcurveto{\pgfxy(15.07,90.94)}{\pgfxy(21.85,95.28)}{\pgfxy(29.18,94.81)}\pgfcurveto{\pgfxy(35.53,94.41)}{\pgfxy(41.09,90.40)}{\pgfxy(43.48,84.50)}\pgfstroke
\pgfmoveto{\pgfxy(12.42,84.09)}\pgflineto{\pgfxy(14.08,86.45)}\pgflineto{\pgfxy(13.18,86.05)}\pgflineto{\pgfxy(12.78,86.95)}\pgflineto{\pgfxy(12.42,84.09)}\pgfclosepath\pgffill
\pgfmoveto{\pgfxy(12.42,84.09)}\pgflineto{\pgfxy(14.08,86.45)}\pgflineto{\pgfxy(13.18,86.05)}\pgflineto{\pgfxy(12.78,86.95)}\pgflineto{\pgfxy(12.42,84.09)}\pgfclosepath\pgfstroke
\color[rgb]{0,0,0}\pgfputat{\pgfxy(28.96,79.46)}{\pgfbox[bottom,left]{\fontsize{5.69}{6.83}\selectfont \makebox[0pt]{target}}}
\pgfputat{\pgfxy(43.42,79.50)}{\pgfbox[bottom,left]{\fontsize{5.69}{6.83}\selectfont \makebox[0pt]{control}}}
\end{pgfpicture}
\end{tabular}
\end{center}
\vspace*{8pt}
  \caption{Logical $\controlled{S}$ gate implemented by the physical $\controlled{S}$ gate.}
  \label{Figure:Controlled-PHASE-gate}
\end{figure}

Hence, using the results of the previous subsection we can conclude that the set $\{C(S),
\iSWAP{\theta}, \PHASE{\theta}\}$ is also universal for quantum computation with respect to our
qubit encoding.

\subsection{Improved constructions for useful quantum gates}

One of the most commonly used quantum gate is the Toffoli (also known as CCNOT) gate. Using the
universal set of $\cnot$ and single qubit gates it can be implemented up to relative phase factor
using the following circuit from \cite{Nielsen-Chuang:2000:QC}:
\begin{figure}
$$
\begin{array}{l}\quad\quad~ \Qcircuit
@C=1.0em @R=1.0em {
\lstick{\ket{c_1}} & \qw & \qw & \qw & \ctrl{2} & \qw & \qw & \qw & \qw\\
\lstick{\ket{c_2}} & \qw & \ctrl{1} & \qw & \qw & \qw & \ctrl{1} & \qw & \qw\\
\lstick{\ket{\psi}} & \gate{R_{\hat{y}}\left(\frac{\pi}{4}\right)} & \targ & \gate{R_{\hat{y}}\left(\frac{\pi}{4}\right)} & \targ & \gate{R_{\hat{y}}\left(-\frac{\pi}{4}\right)} & \targ & \gate{R_{\hat{y}}\left(-\frac{\pi}{4}\right)} & \qw\\
}\end{array}
$$
\vspace*{8pt}
  \caption{Circuit implementing Toffoli gate up to relative phase factor.}
  \label{Figure:Toffoli-gate-standard}
\end{figure}
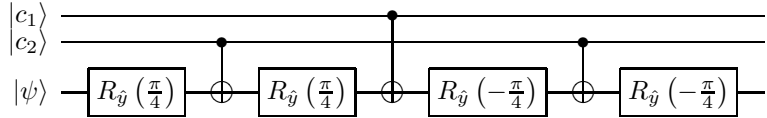

We can construct a more efficient \quotes{low-level} circuit (in a sense that we are transforming
the physical qubits rather than logical ones) using $\cswap$ gates and an extra processing node in
the state $\ket{0}$. The following circuit implements a controlled-controlled-SWAP operation, which
is equivalent to the encoded $\ccnot$ gate:
\begin{figure}
$$
\begin{array}{l}\quad\quad \Qcircuit  @C=1.0em @R=1.0em {
\lstick{\begin{array}{r} \\\left.\ket{c_{1,L}}\right\{\end{array}} & \ctrl{2} & \qw\\
 & \qw & \qw\\
\lstick{\begin{array}{r} \\\left.\ket{c_{2,L}}\right\{\end{array}} & \ctrl{2} & \qw\\
 & \qw & \qw\\
\lstick{\begin{array}{r} \\\left.\ket{\psi_L}\right\{\end{array}} & \qswap \qwx[1]& \qw\\
 & \qswap & \qw\\
\lstick{\ket{0}} & \qw & \qw\\
}\end{array} =
\begin{array}{l}\quad\quad\quad~ \Qcircuit
@C=1.0em @R=1.0em {
\lstick{\begin{array}{r} \\\left.\ket{c_{1,L}}\right\{\end{array}} & \ctrl{2} & \qw & \ctrl{2} & \qw\\
 & \qw & \qw & \qw & \qw\\
\lstick{\begin{array}{r} \\\left.\ket{c_{2,L}}\right\{\end{array}} & \qswap \qwx[4] & \qw & \qswap \qwx[4] & \qw\\
 & \qw & \qw & \qw & \qw\\
\lstick{\begin{array}{r} \\\left.\ket{\psi_L}\right\{\end{array}} & \qw & \qswap \qwx[1] & \qw & \qw\\
 & \qw & \qswap & \qw & \qw\\
\lstick{\ket{0}} & \qswap & \ctrl{-1} & \qswap & \qw\\
}\end{array}
\longrightarrow
\begin{array}{l}\quad\quad\quad \Qcircuit  @C=1.0em @R=1.0em {
\lstick{\ket{c_{1,L}}} & \ctrl{1} & \qw\\
\lstick{\ket{c_{2,L}}} & \ctrl{1} & \qw\\
\lstick{\ket{\psi_L}} & \targ & \qw\\
}\end{array}
$$
\vspace*{8pt}
  \caption{Efficient implementation of the encoded Toffoli gate using the
physical C(C(SWAP)) operation.}
  \label{Figure:Encoded-Toffoli-gate}
\end{figure}
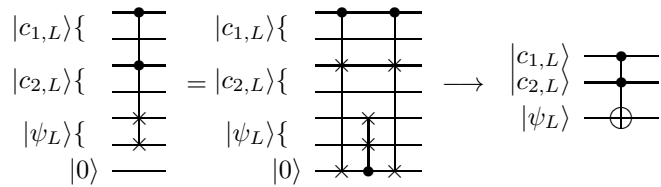

This approach can be generalized to improve constructions of the general $C^t(U)$ gate, defined by
the following equation in \cite{Nielsen-Chuang:2000:QC}:
\begin{equation}
C^t(U)\ket{c_1c_2\ldots c_t}\ket{\psi}=\ket{c_1c_2\ldots c_t}U^{c_1\cdot c_2\cdots c_t}\ket{\psi}.
\end{equation}

The usual construction exploiting ancillary qubits in state $\ket{0}$ is the following
(demonstrated for $t=4$):
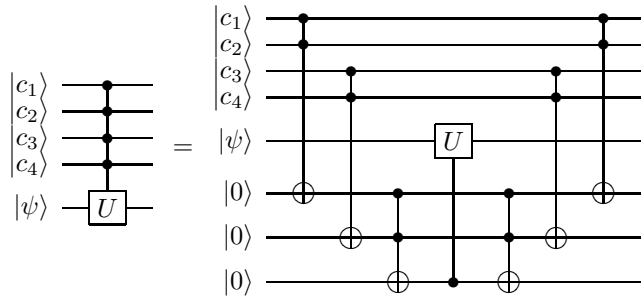
\begin{figure}
$$
\begin{array}{l}\quad\quad \Qcircuit  @C=1.0em @R=1.0em {
\lstick{\ket{c_1}} & \ctrl{1} & \qw\\
\lstick{\ket{c_2}} & \ctrl{1} & \qw\\
\lstick{\ket{c_3}} & \ctrl{1} & \qw\\
\lstick{\ket{c_4}} & \ctrl{1} & \qw\\
\lstick{\ket{\psi}} & \gate{U} & \qw\\
}\end{array} = \begin{array}{l}\quad\quad
\Qcircuit  @C=1.0em @R=1.0em {
\lstick{\ket{c_1}} & \ctrl{1} & \qw & \qw & \qw & \qw & \qw & \ctrl{1} & \qw\\
\lstick{\ket{c_2}} & \ctrl{4} & \qw & \qw & \qw & \qw & \qw & \ctrl{4} & \qw\\
\lstick{\ket{c_3}} & \qw & \ctrl{1} & \qw & \qw & \qw & \ctrl{1} & \qw & \qw\\
\lstick{\ket{c_4}} & \qw & \ctrl{3}& \qw & \qw & \qw & \ctrl{3} & \qw & \qw\\
\lstick{\ket{\psi}} & \qw & \qw & \qw & \gate{U} & \qw & \qw & \qw & \qw\\
\lstick{\ket{0}} & \targ & \qw & \ctrl{1} & \qw & \ctrl{1} & \qw & \targ & \qw\\
\lstick{\ket{0}} & \qw & \targ & \ctrl{1} & \qw & \ctrl{1} & \targ & \qw & \qw\\
\lstick{\ket{0}} & \qw & \qw & \targ & \ctrl{-3} & \targ & \qw & \qw & \qw\\
}
\end{array}
$$
\vspace*{8pt}
  \caption{Implementation of the gate $U$ controlled by four qubits.}
  \label{Figure:Generalized-Controlled-gate-standard}
\end{figure}

This circuit requires $2(t-1)$ Toffoli gates and one controlled-$U$ operation plus $t-1$ pairs of
qubits (initially in the state $\ket{0}$) for temporary storage. As we already know each encoded
Toffoli gate can be implemented using three \quotes{physical} $\cswap$ gates and additional node in
the state $\ket{0}$. Of course, for non-parallel Toffoli gates we can use the same ancillary node,
for it remains in the state $\ket{0}$.

On the other hand, we can use the following scheme for implementing the encoded $C^t(U)$ gate:
\begin{figure}
\[\quad\quad\quad
\Qcircuit  @C=1.0em @R=1.0em {
\lstick{\begin{array}{r} \\\left.\ket{c_{1,L}}\right\{\end{array}} & \ctrl{2} & \qw & \qw & \qw & \qw & \qw & \ctrl{2} & \qw\\
 & \qw & \qw & \qw & \qw & \qw & \qw & \qw & \\
\lstick{\begin{array}{r} \\\left.\ket{c_{2,L}}\right\{\end{array}} & \qswap\qwx[8] & \qw & \qw & \qw & \qw & \qw & \qswap\qwx[8] & \qw\\
 & \qw & \qw & \qw & \qw & \qw & \qw & \qw & \qw\\
\lstick{\begin{array}{r} \\\left.\ket{c_{3,L}}\right\{\end{array}} & \qw & \ctrl{2} & \qw & \qw & \qw & \ctrl{2} & \qw & \qw\\
 & \qw & \qw & \qw & \qw & \qw & \qw & \qw & \qw\\
\lstick{\begin{array}{r} \\\left.\ket{c_{4,L}}\right\{\end{array}} & \qw & \qswap\qwx[5]& \qw & \qw & \qw & \qswap\qwx[5] & \qw & \qw\\
 & \qw & \qw & \qw & \qw & \qw & \qw & \qw & \qw\\
\lstick{\begin{array}{r}\\\\\left.\begin{array}{r}\\ \ket{\psi_L}\end{array}\right\{\end{array}} & \qw & \qw & \qw & \multigate{1}{R} & \qw & \qw & \qw & \qw\\
 & \qw & \qw & \qw & \ghost{R} & \qw & \qw & \qw & \qw\\
\lstick{\ket{0}} & \qswap & \qw & \ctrl{1} & \qw & \ctrl{1} & \qw & \qswap & \qw\\
\lstick{\ket{0}} & \qw & \qswap & \qswap\qwx[1] & \qw & \qswap\qwx[1] & \qswap & \qw & \qw\\
\lstick{\ket{0}} & \qw & \qw & \qswap & \ctrl{-3} & \qswap & \qw & \qw & \qw\\
}
\]
\vspace*{8pt}
  \caption{Improved implementation of the gate $U$ controlled by four qubits.}
  \label{Figure:Generalized-Controlled-gate-efficient}
\end{figure}
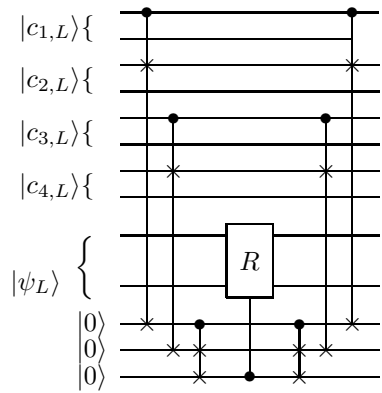

Here we have made use of $2(t-1)$ $\cswap$s (which are elementary gates in our model) and $t-1$
ancillary processing nodes instead of pairs of qubits.

\subsection{Physical realization of swapping gates}

\begin{figure}
  \centerline{\includegraphics[scale=1]{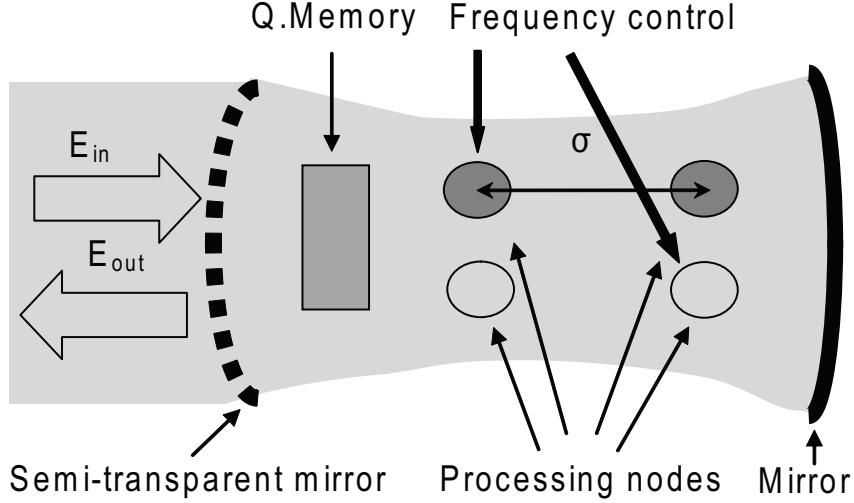}}
\vspace*{8pt}
  \caption{Scheme of quantum computer based on multi-atomic ensembles in single
mode QED cavity coupled with external flying photon qubits. Quantum memory
is used for storage of the photon qubits. The qubits are transferred in two
pairs of processing nodes (blue and green pairs) for realization of single
and two qubit gates. One pair of nodes (with the same color) is used for
encoding of single qubit.}
  \label{QuantumComputer}
\end{figure}

Below we describe the realization of swapping gates controlled by a photon qubit. Let's consider
two atomic ensembles situating in two separate nodes in a common resonator (Fig.
\ref{QuantumComputer}). We can introduce signal (denoted by E$_{in}$ and E$_{out})$ and control
(E$_{c})$ photons through a beam splitter into the system. The photons are stored for a processing
time in quantum memory situating also in common
resonator \cite{Moiseev:2010:quantum-memory,Moiseev-Andrianov:2010:Multi-qubit-QC}. After storage of
the photons, we raise reflectivity of input-output mirror in order to make resonator perfect.
First, signal photon is transferred from quantum memory to two processing nodes that leads to the
following quantum state of the nodes $\ket{\psi_L} =\alpha \ket{0_L} +\beta \ket{1_L}$. Then the
frequency of atomic transitions in processing nodes is tuned out of resonance with the cavity mode.
We release a control photon from the quantum memory and detune the memory from resonance with the
released photon. In that case, the control photon can not be absorbed by the memory and processing
nodes or released from cavity during its lifetime in the cavity.

Such a system of two nodes and control photon in the cavity is described by
the Hamiltonian $H=H_0 +H_1 $, where $H_0 =H_a +H_r $ is a main part and
$H_1 =H_{r-a} =H_{r-a}^{\left( 1 \right)} +H_{r-a}^{\left( 2 \right)} $ is a
perturbation part $H_1 =H_{r-a} =H_{r-a}^{\left( 1 \right)} +H_{r-a}^{\left(
2 \right)} $, $H_a =H_{a_1 } +H_{a_2 } $ is Hamiltonian of atoms in the
nodes 1 and 2. With that, $H_{a_1 } =\hbar \omega _0 \sum\limits_{j_1 }
{S_{j_1 }^z } $ and $H_{a_2 } =\hbar \omega _0 \sum\limits_{j_2 } {S_{j_2
}^z } $ where $\omega _0 $ is the frequency of working transitions in atoms,
$S_{j_1 }^z $ and $S_{j_2 }^z $ operators of effective spin
$\raise.5ex\hbox{$\scriptstyle 1$}\kern-.1em/
\kern-.15em\lower.25ex\hbox{$\scriptstyle 2$} $ z-projection in two-level
model for atoms in sites $j_{1}$ and $j_{2}$ of nodes 1 and 2; $H_r =\hbar
\omega _{k_0 } a_{k_0 \sigma }^+ a_{k_0 \sigma } $, where $\omega _{k_0 } $
is frequency of photons with wave vector $k_0 $, $a_{k_0 \sigma }^+ $ and
$a_{k_0 \sigma } $ are creation and annihilation operators for photons with
wave vector $k_0 $ and polarizations $\sigma $.

We have the following expressions $H_{r-a}^{\left( \sigma \right)}
=H_{r-a}^{\left( {\sigma 1} \right)} +H_{r-a}^{\left( {\sigma 2} \right)} $
for interaction of atoms with photons of polarization $\sigma $ in nodes 1
and 2, where
\begin{equation}
\label{eq8}
H_{r-a}^{\left( {\sigma 1} \right)} =\sum\limits_{j_1 } {\left( {g_{k_0
\sigma }^{\left( 1 \right)} e^{i\vec {k}_0 \vec {r}_{j_1 } }S_{j_1 }^+
a_{k_0 \sigma } +g_{k_0 \sigma }^{\left( 1 \right)\ast } e^{-i\vec {k}_0
\vec {r}_{j_1 } }S_{j_1 }^- a_{k_0 \sigma }^+ } \right)} ,
\end{equation}
\begin{equation}
\label{eq9}
H_{r-a}^{\left( {\sigma 2} \right)} =\sum\limits_{j_2 } {\left( {g_{k_0
\sigma }^{\left( 2 \right)} e^{i\vec {k}_0 \vec {r}_{j_2 } }S_{j_2 }^+
a_{k_0 \sigma } +g_{k_0 \sigma }^{\left( 2 \right)\ast } e^{-i\vec {k}_0
\vec {r}_{j_2 } }S_{j_2 }^- a_{k_0 \sigma }^+ } \right)} ,
\end{equation}
where $g_{k_0 \sigma }^{\left( \alpha \right)} $ are interaction constants,
$S_{j_2 }^+ $ are raising and lowering operators for spin
$\raise.5ex\hbox{$\scriptstyle 1$}\kern-.1em/
\kern-.15em\lower.25ex\hbox{$\scriptstyle 2$} $ in two level model, $\vec
{r}_{j_\alpha } $ are radius vectors for atoms in sites $j_\alpha $ of nodes
$\alpha =1,2$.

We perform unitary transformation \cite{Imamoglu:1999:quantum-dot-spins} of Hamiltonian $H_s
=e^{-s}He^s$ that yields the following result in the second order on small perturbation:
\begin{equation}
\label{eq10}
H_s =H_0 +\frac{1}{2}\left[ {H_1 ,s} \right],
\end{equation}
when relation $H_1 +\left[ {H_0 ,s} \right]=0$ is valid. Using relation (\ref{eq10})
we find $s=s_1 +s_2 $ where
\begin{equation}
\label{eq11}
s_1 =\sum\limits_{j_1 } {\left( {\alpha _1 g_{k_0 \sigma }^{\left( 1
\right)} e^{i\vec {k}_0 \vec {r}_{j_1 } }S_{j_1 }^+ a_{k_0 \sigma } +\beta
_1 g_{k_0 \sigma }^{\left( 1 \right)\ast } e^{-i\vec {k}_0 \vec {r}_{j_1 }
}S_{j_1 }^- a_{k_0 \sigma }^+ } \right)} ,
\end{equation}
\begin{equation}
\label{eq12}
s_2 =\sum\limits_{j_2 } {\left( {\alpha _2 g_{k_0 \sigma }^{\left( 2
\right)} e^{i\vec {k}_0 \vec {r}_{j_2 } }S_{j_2 }^+ a_{k_0 \sigma } +\beta
_2 g_{k_0 \sigma }^{\left( 2 \right)\ast } e^{-i\vec {k}_0 \vec {r}_{j_2 }
}S_{j_2 }^- a_{k_0 \sigma }^+ } \right)} ,
\end{equation}
where $\alpha _{1,2} =-\beta _{1,2} =-\hbar ^{-1}/\left( {\omega _0 -\omega
_{k_0 } } \right)=-\hbar ^{-1}/\Delta $.

Substituting (\ref{eq11}), (\ref{eq12}) into (\ref{eq10}), we get
\begin{equation}
\label{eq13}
\begin{array}{l}
 H_s =\hbar \omega _{k_0 } a_{k_0 \sigma }^+ a_{k_0 \sigma }
+\sum\limits_m^{1,2} {\sum\limits_{j_m } {\hbar \omega _m S_{j_m }^z } }
+2\sum\limits_m^{1,2} {\sum\limits_{j_m }^{N_m } {\frac{\left| {g_{k_0
}^{\left( m \right)} } \right|^2}{\hbar \Delta _m }a_{k_0 \sigma }^+ a_{k_0
\sigma } S_{j_m }^z } } +\\\sum\limits_m^{1,2} {\sum\limits_{i_m j_m }^{N_m
N_m } {\frac{\left| {g_{k_0 }^{\left( m \right)} } \right|^2}{\hbar \Delta
_m }S_{i_m }^+ S_{j_m }^- +} }
 \frac{1}{2\hbar }\left( {\frac{1}{\Delta _1 }+\frac{1}{\Delta _2 }}
\right)\sum\limits_{j_1 j_2 }^{N_1 N_2 } {\left( {g_{k_0 \sigma }^{\left( 1
\right)} g_{k_0 \sigma }^{\left( 2 \right)\ast } e^{i\vec {k}_0 \vec
{r}_{j_1 j_2 } }S_{j_1 }^+ S_{j_2 }^- +h.c. } \right)} . \\
 \end{array}
\end{equation}
The first two terms are unchanged energy of photons, the third term is
unchanged energy of atoms in nodes 1 and 2, the forth and the fifths terms
are atomic energy shifts due to photons with polarization $\sigma $, the
sixes and sevens terms are atomic intra-node swap energies, the eights term
is atomic inter-node swap energy and the nines term is atomic mediated
polarization swapping energy of photons.

Using Hamiltonian (\ref{eq13}) and basic lowest atomic states $\psi _1 =\left| 0
\right\rangle _1 \left| 0 \right\rangle _2 $, $\psi _2 =\left| 1
\right\rangle _1 \left| 0 \right\rangle _2 $, $\psi _3 =\left| 0
\right\rangle _1 \left| 1 \right\rangle _2 $, $\psi _4 =\left| 1
\right\rangle _1 \left| 1 \right\rangle _2 $, $\psi _5 =\left| 2
\right\rangle _1 \left| 0 \right\rangle _2 $ and $\psi _6 =\left| 0
\right\rangle _1 \left| 2 \right\rangle _2 $, we get the following wave
function of the atoms and control field excited in the Fock state with $n$
photons $\ket{n}$
\begin{equation}
\label{eq14}
\ket{\psi\left({n,t}\right)} = \left\{ \sum\limits_{j=1}^6 c_j^{\left( n
\right)} \left( t \right)\psi_j
\right\} \ket{n} ,
\end{equation}
that leads to the following Schr\"{o}dinger equation
\begin{equation}
\label{eq15}
\begin{array}{l}
 \frac{d \langle {n} \left|\psi (n,t)\right\rangle }{dt}=-\frac{i}{\hbar }{\langle {n}\left| H_s \psi (n,t) \right\rangle}=
+\frac{i}{2} {\left\{ { {N_1 \omega _1 +N_2 \omega _2 +2n\left( {N_1
\Omega _1 +N_2 \Omega _2 } \right)}  }
\right\}} c_1^{\left( n \right)} \psi _1 + \\
 +i\left\{ {\left( {\frac{N_1 }{2}-1} \right)\left( {\omega _1 +2n\Omega _1
} \right)+\frac{N_2 }{2}\left( {\omega _2 +2n\Omega _2 } \right)-N_1 \Omega
_1 } \right\}c_2^{\left( n \right)} \psi _2 -i\sqrt {N_1 N_2 } \Omega _s
c_3^{\left( n \right)} \psi _2 + \\
 +i\left\{ {\frac{N_1 }{2}\left( {\omega _1 +2n\Omega _1 } \right)+\left(
{\frac{N_2 }{2}-1} \right)\left( {\omega _2 +2n\Omega _2 } \right)-N_2
\Omega _2 } \right\}c_3^{\left( n \right)} \psi _3 -i\sqrt {N_1 N_2 } \Omega
_s c_2^{\left( n \right)} \psi _3 \\
 +i\left\{ {\left( {\frac{N_1 }{2}-1} \right)\left( {\omega _1 +2n\Omega _1
} \right)+\left( {\frac{N_2 }{2}-1} \right)\left( {\omega _2 +2n\Omega _2 }
\right)-\left( {N_1 \Omega _1 +N_2 \Omega _2 } \right)} \right\}c_4^{\left(
n \right)} \psi _4 - \\
 -i\left\{ {\Omega _s^\ast \sqrt {2N_2 \left( {N_1 -1} \right)} c_5^{\left(
n \right)} +\Omega _s \sqrt {2N_1 \left( {N_2 -1} \right)} c_6^{\left( n
\right)} } \right\}\psi _4 + \\
 +i\left\{\left( {\frac{N_1 }{2}-2} \right)\left( {\omega _1 +2n\Omega _1 }
\right)c_5^{\left( n \right)}  -\Omega _s \sqrt {2N_2 \left( {N_1
-1} \right)} c_4^{\left( n \right)} \right\}\psi _5 + \\
 +i\left\{\left( {\frac{N_2 }{2}-2} \right)\left( {\omega _2 +2n\Omega _2 }
\right)c_6^{\left( n \right)}  -\Omega _s \sqrt {2N_1 \left( {N_2
-1} \right)} c_4^{\left( n \right)}\right\} \psi _6 ,
 \end{array}
\end{equation}
where $\Omega _1 =\frac{\left| {g_{k_0 }^{\left( 1 \right)} }
\right|^2}{\hbar ^2\Delta _1 }$, $\Omega _2 =\frac{\left| {g_{k_0 }^{\left(
2 \right)} } \right|^2}{\hbar ^2\Delta _2 }$ and $\Omega _s =\frac{g_{k_0
}^{\left( 1 \right)} g_{k_0 }^{\left( 2 \right)\ast } }{2\hbar ^2}\left(
{\frac{1}{\Delta _1 }+\frac{1}{\Delta _2 }} \right)$.

Below we are interested only in the dynamics of the amplitudes $c_2^{(n)} $
and $c_3^{(n)} $ controlled by the presence or absence of the cavity photon
state $\vert $n$>$
\begin{equation}
\label{eq16}
\frac{dc_2^{(n)} }{dt}=\frac{i}{\hbar }E_2 (n)c_2^{(n)} -i\sqrt {N_1 N_2 }
\Omega _s c_3^{(n)} ,
\end{equation}
\begin{equation}
\label{eq17}
\frac{dc_3^{(n)} }{dt}=\frac{i}{\hbar }E_3 (n)c_3^{(n)} -i\sqrt {N_1 N_2 }
\Omega _s c_2^{(n)} ,
\end{equation}
where $E_{2,3} (n)=\bar {E}(n)+\delta E_{2,3} (n)$, $\bar
{E}(n)=\textstyle{1 \over 2}\hbar \left\{ {N_1 \left( {\omega _1 +2n\Omega
_1 } \right)+N_2 (\omega _2 +2n\Omega _2 )} \right\}$, $\delta E_2
(n)=-\hbar \left\{ {\left( {\omega _1 +2n\Omega _1 } \right)+N_1 \Omega _1 }
\right\}$, $\delta E_3 (n)=-\hbar \left\{ {\left( {\omega _2 +2n\Omega _2 }
\right)+N_2 \Omega _2 } \right\}$ with a solution
\begin{equation}
\label{eq18}
\begin{array}{l}
c_2^{(n)} =e^{i\Delta (n)t}\{A_1 (n)e^{iS(n)t}+A_2 (n)e^{-iS(n)t}\},\\
c_3^{(n)} =e^{i\Delta (n)t}\{B_1 (n)e^{iS(n)t}+B_2 (n)e^{-iS(n)t}\},
\end{array}
\end{equation}
where $\Delta (n)=\textstyle{1 \over {2\hbar }}[E_2 (n)+E_3 (n)]$,
$S(n)=\sqrt {\left( {\left\{ {\delta E_2 (n)-\delta E_3 (n)} \right.}
\right)^2+4N_1 N_2 \hbar ^2\Omega _s^2 } $.

By taking into account the initial conditions $c_2^{(n)} =1$ and $c_3^{(n)}
=0$, we get $B_1 (n)=-B_2 (n)=-\frac{\sqrt {N_1 N_2 } \Omega _s }{2S(n)}$,
(24)
\begin{equation}
\label{eq19}
c_3^{(n)} (t)=2ie^{i\Delta (n)t}B_1 (n)\sin [S(n)t],
\end{equation}
and
\begin{equation}
\label{eq20}
A_1 (n)=\frac{1}{2}\left( {1+\frac{\left( {\delta E_2 (n)-\delta E_3 (n)}
\right)}{S(n)}} \right),
\end{equation}
\begin{equation}
\label{eq21}
A_2 (n)=\frac{1}{2}\left( {1-\frac{\left( {\delta E_2 (n)-\delta E_3 (n)}
\right)}{S(n)}} \right).
\end{equation}
For equal resonant frequencies of the two nodes in vacuum cavity mode state
$\omega _1 +N_1 \Omega _1 =\omega _2 +N_2 \Omega _2 $, $\delta E_2
(n)-\delta E_3 (n)=2n\hbar \left( {\Omega _1 -\Omega _2 } \right)$,
$S(n)=\sqrt {n^2\left( {\Omega _1 -\Omega _2 } \right)^2+N_1 N_2 \Omega _s^2
} $. Expression (24) simplifies to the following
\begin{equation}
\label{eq22}
B_1 (n)=-B_2 (n)=-\frac{S(0)}{2S(n)}.
\end{equation}
and expressions (\ref{eq19}), (\ref{eq20}) simplify to
\begin{equation}
\label{eq23}
A_1 (n)=\frac{1}{2}\left( {1+\frac{n\left( {\Omega _2 -\Omega _1 }
\right)}{S(n)}} \right),
\end{equation}
\begin{equation}
\label{eq24}
A_2 (n)=\frac{1}{2}\left( {1-\frac{n\left( {\Omega _2 -\Omega _1 }
\right)}{S(n)}} \right).
\end{equation}
Below we are interested in the quantum dynamics of the processing nodes controlled by a single
photon field where the obtained solutions are

\textbf{1. CSWAP gate based on dynamical elimination of $c_3^{(1)} $}

\noindent $\ket{n=0}$ -- photon state:
\begin{equation}
\label{eq25}
c_2^{(0)} (t)=e^{i\Delta (0)T}\cos \{S(0)t\},
\end{equation}
\begin{equation}
\label{eq26}
c_3^{(0)} (t)=-ie^{i\Delta (0)t}\sin \{S(0)t\}.
\end{equation}
$\ket{n=1}$ -- photon state:
\begin{equation}
\label{eq27}
c_2^{(1)} (t)=e^{i\Delta (1)t}\{\cos [S(1)t]+i\frac{\left( {\Omega _2
-\Omega _1 } \right)}{S(1)}\sin [S(1)t]\},
\end{equation}
\begin{equation}
\label{eq28}
c_3^{(1)} (t)=-ie^{i\Delta (1)t}\frac{S(0)}{S(1)}\sin [S(1)t],
\end{equation}
where $S(0)=\sqrt {N_1 N_2 } \Omega _s $, $S(1)=\sqrt {\left( {\Omega _2
-\Omega _1 } \right)^2+N_1 N_2 \Omega _s^2 } $.

We can realize the $\cswap$ gate by using the quantum evolution of the
processing gates during fixed temporal interval t=t$_{\cswap}$=T where
$S(0)T=\pi /2+n\pi $ and $S(\ref{eq1})T=\pi +m\pi $ which leads to the following
relation $S(\ref{eq1})=\frac{1+m}{(1/4+n)}S(0)$ ($m=0,1,2,{\ldots}$; $n=1,2,{\ldots}$).
By assuming $m=0$ and $n=1$, we get $\vert \Omega _1 -\Omega _2 \vert
=\textstyle{1 \over 2} \sqrt {5 N_1 N_2} \Omega
_s $. At equal values of atom numbers in the nodes $N_1 =N_2 =N$ and
frequency offsets $\Delta _1 =\Delta _2 =\Delta $, we have $\Omega _s^2
=\Omega _1 \Omega _2 $ and
$\Omega _1 =1.25 N^2 \Omega _2[\sqrt {\textstyle{1 \over 4}+(\textstyle{16 \over 25}) N^{-2}}+\textstyle{1 \over 2}] $.

\textbf{2. $\cswap$ gate based on strong blockade of $c_3^{(1)} $}

Here, we see that if $n=1$ and $\vert \Omega _2 -\Omega _1 \vert >>\Omega _s
\sqrt {N_1 N_2 } $ we have $B_1 =-B_2 =c_3 \cong 0$, $A_1 =1$ and $A_2 \cong
0$. If $n=0$ we have $B_1 =-B_2 =-\frac{1}{2}$ and $A_1 =A_2 =\frac{1}{2}$.
In the first case no swap occurs and in the second case we have swapping
solution
\begin{equation}
\label{eq29}
c_3^{(1)} \cong -ie^{\frac{i}{2\hbar }\left( {E_2 +E_3 } \right)t}\sin \left( {\sqrt
{N_1 N_2 } \Omega _s t} \right),
\end{equation}
and
\begin{equation}
\label{eq30}
c_2^{(1)} \cong e^{\frac{i}{2\hbar }\left( {E_2 +E_3 } \right)t}\cos \left( {\sqrt {N_1
N_2 } \Omega _s t} \right).
\end{equation}
State $\psi _2 $ is converted into state $\psi _3 $ on time interval $t_{\cswap} =\pi /2\Omega _S
\sqrt {N_1 N_2 } $. Thus, we have swap gate controlled by photon state ($\cswap$ gate). By
substituting the blockade condition we find that $\cswap$ gate operates with single atomic rate
$t_{\cswap} =\pi /{(2\Omega _{S} N)} =t_{\cswap} = \pi /{(2 \sqrt {\Omega_1 \Omega_2} N)} \approx
\pi /{(2\Omega_1 )}$. While the SWAP gate does operate $N$ times faster in multi-atomic case in
accordance with Eqs.~(\ref{eq29}), (\ref{eq30}). Therefore,
 it is reasonable to use single atoms (molecules) or quantum dots with a large dipole moment for $\cswap$ operation,
 while using multi-atomic ensembles for simple SWAP operation. They can yield CNOT and single qubit gates with the qubit encoding
from subsection \ref{section:Universality}.

According to subsection \ref{section:Universality}, we need the have a swap operation in the pair
of atomic ensembles (two green nodes in Fig. \ref{QuantumComputer}) controlled by another pair of
atomic ensembles (two blue nodes in Fig. \ref{QuantumComputer}). This can be achieved by releases
photon from one node of blue node pair by equalizing the node frequency with the frequency QED
cavity mode and switching of the node frequency from the resonance in a proper moment of time.
After the $\cswap$ gate realization, we can return the photon to the chosen blue node by equalizing
the node frequency with the cavity node. This $\cswap$ gate is equivalent to $\cnot$ gate with
qubit encoding introduced in section \ref{section:Universality}. Such gates and other necessary
gates can be incorporated in common cavity in a quantity that is needed for implementation of one
or another quantum algorithm. All of them can be initiated through a single quantum memory that is
essentially multi-qubit in photon echo approach \cite{Moiseev:2010:quantum-memory}.

\section{Conclusion}

Summarizing, we have proposed an approach for constructing encoded universal quantum
computations based on swapping operations. The main idea of the proposed set of quantum gates
is encoding logical qubits by two physical qubits. It allows to \emph{explicitly}
implement any encoded single-qubit gate by 3 elementary operations and to perform encoded controlled-NOT gate
\emph{by a single} $\cswap$ operation on pairs of atomic ensembles.
We have shown that our basis of quantum gates is efficient (in terms of the number of elementary
gates) for implementing complex controlled operations, which are at the core of many efficient
quantum algorithms.

This approach considerably simplifies physical implementation of quantum computer on
multi-atomic ensembles in the QED resonator at the price of doubling the number of qubits for
computation.
Besides, it permits to avoid the necessity of implementing blockade of excess states in the multi-atomic
ensembles.
The physical implementation of the basic gates is sufficiently robust and
provides fast single qubit operations based on multi-atomic ensembles.

 Note also, that using of two atomic ensembles for encoding of a single qubit
state will be convenient for the quantum computer interface with the external quantum information
carried by using photon polarization qubits since the two polarization components of the photon
qubit can be coupled directly with the relevant pair of the atomic ensemble state.

\section*{Acknowledgments}

Work was in part supported by the Russian Foundation for Basic Research under the grants
09-01-97004, 10-02-01348, 11-07-00465.

\bibliography{references,physics}

\begin{thebibliography}{10}

\bibitem{Nielsen-Chuang:2000:QC}
Michael~A. Nielsen and Isaac~L. Chuang.
\newblock {\em Quantum Computation and Quantum Information}.
\newblock Cambridge University Press, 1 edition, October 2000.

\bibitem{Nakahara-Ohmi:2008:QC}
Mikio Nakahara and Tetsuo Ohmi.
\newblock {\em Quantum Computing: From Linear Algebra to Physical
  Realizations}.
\newblock CRC Press, Taylor \& Francis, 2008.

\bibitem{Kok:2007:OpticalQC}
Pieter Kok, W.~J. Munro, Kae Nemoto, T.~C. Ralph, Jonathan~P. Dowling, and
  G.~J. Milburn.
\newblock Linear optical quantum computing with photonic qubits.
\newblock {\em Rev. Mod. Phys.}, 79(1):135--174, Jan 2007.

\bibitem{Ladd:2010:QC}
T.~D. Ladd, F.~Jelezko, R.~Laflamme, Y.~Nakamura, C.~Monroe, and J.~L.
  O/'Brien.
\newblock Quantum computers.
\newblock {\em Nature}, 464(7285):45--53, March 2010.

\bibitem{Deutsch:1989:QuantumCircuits}
David Deutsch.
\newblock Quantum computational networks.
\newblock {\em Royal Society of London Proceedings Series A}, 425:73--90, sep
  1989.

\bibitem{DiVincenzo:1995:universal}
David~P. DiVincenzo.
\newblock Two-bit gates are universal for quantum computation.
\newblock {\em Phys. Rev. A}, 51(2):1015--1022, Feb 1995.

\bibitem{Deutsch:1995:Universality}
David Deutsch, Adriano Barenco, and Artur Ekert.
\newblock {Universality in Quantum Computation}.
\newblock {\em Proceedings of the Royal Society of London. Series A:
  Mathematical and Physical Sciences}, 449(1937):669--677, 1995.

\bibitem{Boykin:2000:quantum-basis}
P.~Oscar Boykin, Tal Mor, Matthew Pulver, Vwani Roychowdhury, and Farrokh
  Vatan.
\newblock A new universal and fault-tolerant quantum basis.
\newblock {\em Information Processing Letters}, 75(3):101--107, 2000.

\bibitem{DiVincenzo:2000:Exchange-Interaction}
David~P. DiVincenzo, Dave Bacon, Julia Kempe, Guido Burkard, and K.~Birgitta
  Whaley.
\newblock Universal quantum computation with the exchange interaction.
\newblock {\em Nature}, 408:339--342, 2000.

\bibitem{Bacon:2000:Fault-Tolerant}
Dave Bacon, Julia Kempe, Daniel~A. Lidar, and K.~Birgitta Whaley.
\newblock Universal fault-tolerant quantum computation on decoherence-free
  subspaces.
\newblock {\em Phys. Rev. Lett.}, 85(8):1758--1761, Aug 2000.

\bibitem{Kempe:2001:decoherence-free-computation}
Julia Kempe, Dave Bacon, Daniel~A. Lidar, and K.~Birgitta Whaley.
\newblock Theory of decoherence-free fault-tolerant universal quantum
  computation.
\newblock {\em Phys. Rev. A}, 63(4):042307, Mar 2001.

\bibitem{Kempe:2001:Encoded-Universality}
Julia Kempe, Dave Bacon, David~P. DiVincenzo, and K.~Birgitta Whaley.
\newblock Encoded universality from a single physical interaction.
\newblock In R.~Clark, G.~Milburn, R.~Hughes, A.~Imamoglu, and P.~Delsing,
  editors, {\em Quantum Computation and Information}, volume~1, pages 33--55.
  Rinton Press, New Jersey, 2001.

\bibitem{Kempe:2002:Exact-gate-sequences}
Julia Kempe and K.~Birgitta Whaley.
\newblock Exact gate sequences for universal quantum computation using the $xy$
  interaction alone.
\newblock {\em Phys. Rev. A}, 65(5):052330, May 2002.

\bibitem{Levy:2002:SpinPairs}
Jeremy Levy.
\newblock Universal quantum computation with spin-$1/2$ pairs and heisenberg
  exchange.
\newblock {\em Phys. Rev. Lett.}, 89(14):147902, Sep 2002.

\bibitem{Brion:2007:Ensembles}
E.~Brion, K.~M\o{}lmer, and M.~Saffman.
\newblock Quantum computing with collective ensembles of multilevel systems.
\newblock {\em Phys. Rev. Lett.}, 99(26):260501, Dec 2007.

\bibitem{Saffman-Molmer:2008:RydbergQC}
M.~Saffman and K.~M\o{}lmer.
\newblock Scaling the neutral-atom rydberg gate quantum computer by collective
  encoding in holmium atoms.
\newblock {\em Phys. Rev. A}, 78(1):012336, Jul 2008.

\bibitem{Saffman:2010:RydbergQI}
M.~Saffman, T.~G. Walker, and K.~M\o{}lmer.
\newblock Quantum information with rydberg atoms.
\newblock {\em Rev. Mod. Phys.}, 82(3):2313--2363, Aug 2010.

\bibitem{Shahriar:2007:atomic-ensemblesQCC}
M.~S. Shahriar, G.~S. Pati, and K.~Salit.
\newblock Quantum communication and computing with atomic ensembles using a
  light-shift-imbalance-induced blockade.
\newblock {\em Phys. Rev. A}, 75(2):022323, Feb 2007.

\bibitem{Moiseev:2010:multi-ensembleQC}
Sergey~A. Moiseev, Sergey~N. Andrianov, and Firdus~F. Gubaidullin.
\newblock Solid state multi-ensemble quantum computer in waveguide circuit
  model.
\newblock Technical Report arXiv:1009.5771, Cornell University Library, Sep
  2010.

\bibitem{Andrianov-Moiseev:2011:swapping-gates}
Sergey~N. Andrianov and Sergey~A. Moiseev.
\newblock Fast and robust two- and three-qubit swapping gates on multi-atomic
  ensembles in quantum electrodynamic cavity.
\newblock {\em Electronic Proceedings in Theoretical Computer Science},
  52:13--21, 2011.

\bibitem{Duan-Kimble:2004:PhotonicQC}
L.-M. Duan and H.~J. Kimble.
\newblock Scalable photonic quantum computation through cavity-assisted
  interactions.
\newblock {\em Phys. Rev. Lett.}, 92(12):127902, Mar 2004.

\bibitem{Aoki:2006:strong-coupling}
Takao Aoki, Barak Dayan, E.~Wilcut, W.~P. Bowen, A.~S. Parkins, T.~J.
  Kippenberg, K.~J. Vahala, and H.~J. Kimble.
\newblock {Observation of strong coupling between one atom and a monolithic
  microresonator}.
\newblock {\em Nature}, 443(7112):671--674, 2006.

\bibitem{Majer:2007:CoupledQubits}
J.~Majer, J.~M. Chow, J.~M. Gambetta, Jens Koch, B.~R. Johnson, J.~A. Schreier,
  L.~Frunzio, D.~I. Schuster, A.~A. Houck, A.~Wallraff, A.~Blais, M.~H.
  Devoret, S.~M. Girvin, and R.~J. Schoelkopf.
\newblock {Coupling superconducting qubits via a cavity bus}.
\newblock {\em Nature}, 449(7161):443--447, September 2007.

\bibitem{Palma:1996:quantum-dissipation}
G.~Massimo Palma, Kalle-Antti Suominen, and Artur~K. Ekert.
\newblock Quantum computers and dissipation.
\newblock {\em Proceedings of the Royal Society of London. Series A:
  Mathematical, Physical and Engineering Sciences}, 452(1946):567--584, 1996.

\bibitem{Byrd-Lidar:2002:Problems-of-Decoherence}
Mark~S. Byrd and Daniel~A. Lidar.
\newblock Comprehensive encoding and decoupling solution to problems of
  decoherence and design in solid-state quantum computing.
\newblock {\em Phys. Rev. Lett.}, 89(4):047901, Jul 2002.

\bibitem{ablayev-vasiliev:2009:EPTCS}
Farid Ablayev and Alexander Vasiliev.
\newblock Algorithms for quantum branching programs based on fingerprinting.
\newblock {\em Electronic Proceedings in Theoretical Computer Science},
  9:1--11, 2009.

\bibitem{Moiseev:2010:quantum-memory}
Sergey~A. Moiseev, Sergey~N. Andrianov, and Firdus~F. Gubaidullin.
\newblock Efficient multimode quantum memory based on photon echo in an optimal
  qed cavity.
\newblock {\em Phys. Rev. A}, 82(2):022311, Aug 2010.

\bibitem{Moiseev-Andrianov:2010:Multi-qubit-QC}
Sergey~A. Moiseev and Sergey~N. Andrianov.
\newblock Multi-qubit quantum memory integrated in quantum computer.
\newblock In {\em Proceedings of the 5-th International Scientific school
  ``Science and innovation-2010'', ISS ``SI-2010'': 19 -- 24 July 2010,
  Ioshkar-Ola}, pages 156--164, 2010.

\bibitem{Imamoglu:1999:quantum-dot-spins}
A.~Imamoglu, D.~D. Awschalom, G.~Burkard, D.~P. DiVincenzo, D.~Loss,
  M.~Sherwin, and A.~Small.
\newblock Quantum information processing using quantum dot spins and
  cavity-qed.
\newblock {\em Physical Review Letters}, 83(20):4204--4207, 1999.

\end{thebibliography}

\end{document}